\documentclass[twocolumn]{aastex631}
\usepackage{multirow}
\usepackage{xspace}
\usepackage{gensymb}

\renewcommand{\arraystretch}{2}
\usepackage{graphicx}
\usepackage{amsmath}
\usepackage[caption=false]{subfig}

\newcommand{\Msun}{\,{\rm M_\odot}}

\newcommand{\Mblack}{M_\bullet}
\newcommand{\Mstar}{M_\star}

\newcommand{\mmstar}{$\Mblack-\Mstar$\xspace}

\begin{document}

\title{Exploring the AGN Fraction of a Sample of JWST's Little Red Dots at $4 < z < 8$: \\ Overmassive Black Holes Are Strongly Favored}

\author[0009-0004-9516-9593]{Emmanuel Durodola}
\affiliation{Department of Physics \& Astronomy, Dartmouth College, Hanover, NH 03755, USA}

\author[0000-0001-9879-7780]{Fabio Pacucci}
\affiliation{Center for Astrophysics $\vert$ Harvard \& Smithsonian, Cambridge, MA 02138, USA}
\affiliation{Black Hole Initiative, Harvard University, Cambridge, MA 02138, USA}

\author[0000-0003-1468-9526]{Ryan C. Hickox}
\affiliation{Department of Physics \& Astronomy, Dartmouth College, Hanover, NH 03755, USA}




\begin{abstract}

JWST is revolutionizing our view of the early Universe by pushing the boundaries of detectable galaxies and black holes in redshift (upward) and mass (downward). The Little Red Dots (LRDs), detected by several surveys at $z > 4$, present a significant interpretational challenge, as their Spectral Energy Distributions (SED) can mimic both AGN and stellar population templates. This study analyzes 19 LRDs from the JADES survey, utilizing NIRCam and MIRI photometry. By performing SED fitting across a vast parameter space, we explore a broad range of AGN fractions, defined as the ratio of the monochromatic luminosities (AGN, galaxy, and dust) over a specified wavelength range, 0.4 - 0.7 $\mu m$ rest-frame. We find that 17 of the 19 LRDs investigated are consistent with having significant AGN contributions, with best-fitting AGN fractions ranging between 20\% and 70\%, while one galaxy shows a low AGN contribution (2\%) and another appears to be purely star-forming. Moreover, assuming these LRDs do indeed host AGN, we can place limits on their black hole masses using the inferred AGN bolometric luminosities and adopting the Eddington limit. We find that, independent of the specific AGN fraction adopted, the LRDs' black holes are significantly overmassive relative to their host galaxies---by $\sim 1$ dex, and up to $\sim 4$ dex in the most extreme cases---compared to the local $M_{\bullet} - M_{\star}$ relation. 
The presence of overmassive black holes in the high-$z$ Universe may provide the strongest evidence yet of heavy black hole seeding occurring during the cosmic dark ages.

\end{abstract}


\section{Introduction} 
\label{sec:intro}

JWST provides an unprecedented view of the early Universe. Unlike earlier telescopes that observed massive quasars at $z > 6$ \citep{Fan_2001,Fan_2003, Fan_2006,cool_06,chris_09, wang_millimeter_2007, Willott_2011, Mortlock_2011, Wu_2015, Banados_2018, Pacucci_2019_lensed, Fan_2022_review}, where the total spectral energy distribution (SED) is dominated by the emission of the central supermassive black hole (SMBH, with mass $\Mblack > 10^9 \Msun$), JWST can often capture the light from the host's stellar component (see, e.g., \citealt{ding_detection_2023}), especially in the case of lower-mass (i.e., $\Mblack \sim 10^{6-8} \Msun$), lower-luminosity black holes (i.e., $L_{\rm bol} \sim 10^{44-46} \, \rm erg \, s^{-1}$). This extraordinary sensitivity has led to the detection of new populations of SMBHs in the high-$z$ Universe, farther and with lighter and more compact hosts (e.g., GN-z11 \citealt{Maiolino_2023, Baggen_2023, greene_uncover_2023}). 

Observations of high-$z$ systems play an essential role in our ability to understand the early evolution and co-evolution of black holes and their host galaxies. 
Extensive studies in the local Universe (i.e., $z < 0.1$) have shown the existence of evolutionary relationships between the central SMBHs and their hosts, at least in massive galaxies (see, e.g., \citealt{Ferrarese_Merritt_2000, Gebhardt_2000, kormendy_coevolution_2013}). These observations led to empirical relations between the mass of the central black hole and several properties of the host. For example, the $\Mblack-M_{\star}$ with the stellar mass of the host galaxy, the $\Mblack-\sigma$ with its velocity dispersion, and the $\Mblack-M_{\rm dyn}$ with the dynamical mass of the host  \citep{hu_08,kormendy_coevolution_2013,heckman_14, reines_relations_2015}. While nuances in how these parameters are measured could have some underlying effect on the normalization and scatter in the respective relationships (e.g., \citealt{Shankar_2019}), the overall trend shows an undeniable link between the growth of SMBHs and their host galaxies, at least in the local Universe. This co-evolution is not yet well investigated in the high-$z$ Universe, primarily because the detection of the stellar component of galaxies hosting very massive black holes was rare in the pre-JWST era.

Efforts to quantify these scaling relations using JWST observations at $z > 4$ have revealed a population of central SMBHs that are overmassive compared to the stellar content of their hosts \citep{Maiolino_2023, Maiolino_2023_new, Harikane_2023, Larson_2023, Ubler_2023, Stone_2023, Furtak_2023, Kocevski_2023, Kokorev_2023, Yue_2023, Juodzbalis_2024, Kocevski_2024, Taylor_2024}, but seem to agree with the stellar velocity dispersion and the dynamical mass \citep{Maiolino_2023_new}. More specifically, the $\Mblack-M_{\star}$ relation inferred from $z=4-7$ JWST detections deviates by $>3 \sigma$ from the local relationship: black holes are overmassive by factors of $10\times$ to $100\times$ with respect to local counterparts hosted by a galaxy of the same size \citep{pacucci_jwst_2023}. 

The prevalence of such overmassive black holes brings into question the seeding mechanism responsible for the formation of such high-$z$ systems \citep{pacucci_jwst_2023, Volonteri_2023, Natarajan_2023, Pacucci_2024_evolution}.
Heavy seed scenarios (see, e.g., \citealt{volonteri_08}, and the reviews by \citealt{Woods_2019, inayoshi_2020}) were first introduced to explain the extremely massive SMBHs observed at $z > 6$ (see, e.g., \citealt{Mortlock_2011, Wu_2015, Banados_2018}), which would otherwise lack the necessary time to grow from standard, low-mass seeds \citep{Pacucci_2022_search}. These heavy seeding models refer to various channels that may have shaped the formation of black holes in the early Universe, such as direct collapse black holes \citep{Loeb_Rasio_1994, Bromm_Loeb_2003, Lodato_Natarajan_2006}. Some recent studies (e.g., \citealt{pacucci_jwst_2023, Natarajan_2023, Scoggins_2023, Pacucci_2024_evolution}) suggest that the high-$z$ overmassive black holes observed by JWST may be the direct result of this heavy seeding scenario.

However, determining the location of these high-$z$ systems in the \mmstar plane crucially depends on how the contributions from the stellar and the active galactic nucleus (AGN) components of the SEDs are evaluated, as well as the individual uncertainties in the determination of the stellar mass and the black hole mass (in this regard, see also, e.g., \citealt{Pacucci_2022_dropout}). This study investigates how different assumptions on the AGN/star ratio (i.e., how much these two components generate the observed flux) affect the goodness-of-fit with the observed photometry and the system's location on the \mmstar plane.

In particular, we analyze 19 little red dots (LRDs) first reported in \citealt{barro_extremely_2023} and observed with JWST.
As detailed in Sec. \ref{sec:data}, these EROs are a subset of the more general population of ``Little Red Dots'', or LRDs, a name that was first introduced by \cite{Matthee_2023} and that is now of widespread use.
Since the dataset from \cite{barro_extremely_2023} was published, several other studies reported the detection of similarly red and compact objects, which are ubiquitous in many JWST fields (see, e.g., the extensive catalog provided by \citealt{Kocevski_2024}). In particular, \cite{greene_uncover_2023} argued that most (i.e., $\sim 60\%$) of these sources host a SMBH at their center, based on the detection of broad line regions (BLRs).

The rest of this paper is organized as follows: Section \ref{sec:data} discusses the data used and summarizes the data reduction process. Section \ref{sec:method} details our SED modeling and choices of parameters. In Section 4 and \ref{sec:conclusions}, we discuss results from the SED modeling and its implication within the larger context of galaxy and black hole co-evolution.

\section{Data}
\label{sec:data}

This study is based on the sample of LRDs by \cite{Perez_Gonzalez_2024}, which is currently the deepest publicly available catalog of galaxies observed with the Mid-Infrared instrument (MIRI) on JWST. As discussed in detail later on, this long-wavelength data is essential to constraining the presence of AGN in LRDs. Hence, we focus our efforts on the dataset containing the longest wavelength MIRI observation of LRDs.

\cite{Perez_Gonzalez_2024} reports the discovery of 31 LRDs, whose selection is based on NIRcam colors F277W-F444W $>$ 1 mag, F150W-F200W $<$ 0.5 mag, and F444W $\leq$ 28 mag. These objects are selected from the JWST Advanced Deep Extragalactic Survey (JADES, \citealt{Einsenstein_2023}), with matched Mid-IR observation from the SMILES survey. We select a sub-sample of 19 sources, which are characterized by MIRI detection, within $5 \, \rm \mu m$ and $28 \, \rm \mu m$. We only adopt fluxes in the NIRCam filters (F090W, F115W, F150W, F182M, F200W, F210M, F277W, F335M, F356W, F410M, F430M, F444W, F460M, F480M), and MIRI filters (F560W, F770W, F1000W, F1280W, F1500W, F1800W, F2100W, F2550W). Similar to \citealt{Perez_Gonzalez_2024}, we impose a 5$\sigma$ upper limit for filters where there are non-detections. For the galaxies with non-detections in the short NIRCam wavelengths, we supplement the photometry using 0.25'' aperture HST fluxes and 5$\sigma$ upper limits in both F606W and F814W, as provided in the JADES catalog.

18 out of the 31 galaxies had spectroscopic redshifts from the survey where they were selected. Photometric redshifts were estimated for the remaining galaxies using SED fitting with \texttt{EAZYPY}. The total \cite{Perez_Gonzalez_2024} sample had redshifts between z = 4 - 8.5, with our selected sub-sample spanning a slightly smaller redshift range of z = 4 - 8. 15 of the 18 LRDs with spectroscopic redshifts are included in the analysis presented here.

One of the galaxies selected in our sample, JADES-204851, has been reported to have a broad 2200 km s$^{-1}$ H$\alpha$ component, arising from a 10$^{7.5}$ M$_\odot$ SMBH \citep{Matthee_2023}. For additional details on these LRDs, we refer the reader to \cite{Perez_Gonzalez_2024}.

\section{Methods: SED modeling}
\label{sec:method}

\begin{figure*}
    \centering
    \includegraphics[width = \textwidth]{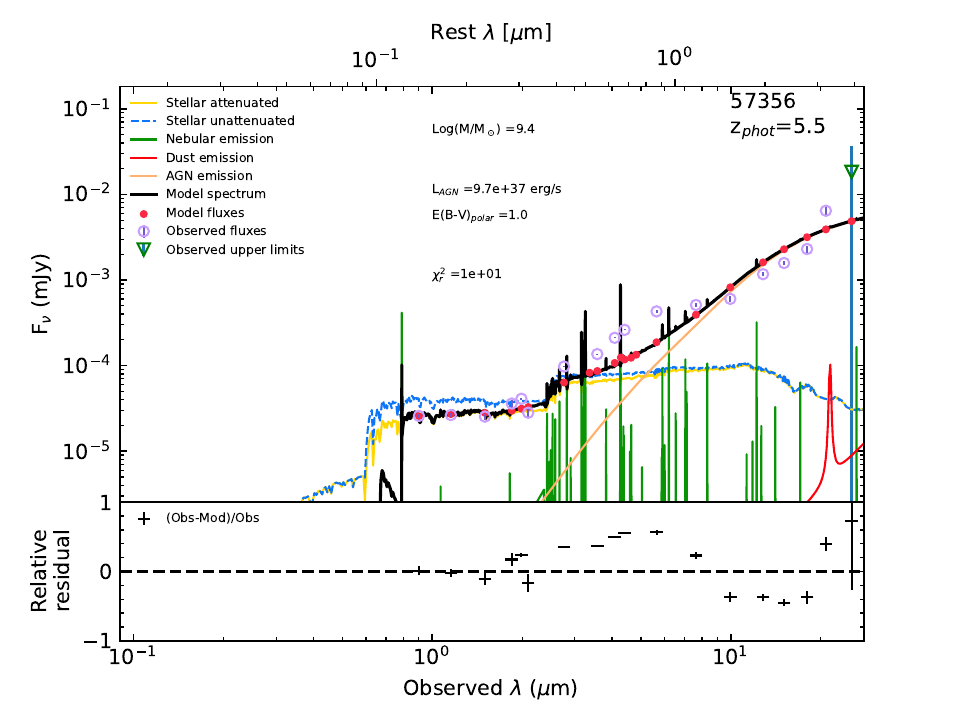}
    \caption{Example of photometric fitting of one of our EROs: JADES-57356. The photometric points of the source and the inferred photometry, based on the components listed, are shown. The bottom panel displays the relative residuals between the observed and inferred photometry.}
    \label{fig:modeling}
\end{figure*}

\begin{table*}[!htb]
\centering
\begin{tabular}{@{}l p{4.5cm} c c @{}}
\hline\hline
\textbf{Module} & \textbf{Parameter} & \textbf{Symbol} & \textbf{Values} \\ \hline

Star Formation History & e-folding time (main)  &  $\tau_{ main}$ &  0.1 - 5 (30 steps) Gyrs \\  

\fontfamily{cmtt}\selectfont{sfhdelayed} &Main stellar age &  $t_{star}$ &  0.1 - 2 (30 steps) Gyrs \\ \hline

Simple stellar population & Initial mass function & ... & Chabrier \\
\fontfamily{cmtt}\selectfont{bc03} & Metallicity & $Z$ & 0.02 \\ \hline

Nebular Emission & Ionization parameter & log $U$ & -2.0 \\
\fontfamily{cmtt}\selectfont{nebular} & Gas Metallicity & $Z_{gas}$ & 0.02 \\ 

\hspace{1cm} & Lyman $\alpha$ escape fraction & $f_{esc}$ & 0.09 \\ \hline

Dust attenuation & Color excess of nebular lines & E(B-V)$_{line}$ & 0.1 \\ \hline

AGN emission & Opening angle & $\theta$ & 40\degree \\
\hspace{1cm} & Viewing angle & $i$ & 30\degree \\
\fontfamily{cmtt}\selectfont{Skirtor 2016} & AGN contribution & frac$_{AGN}$ & 0.02 - 0.2 (step 0.02), 0.2 - 0.9 (step 0.1), 0.99 \\
\hspace{1cm} & Wavelength range for frac$_{AGN}$ & $\lambda_{AGN}$ & 0.4 - 0.7 $\mu$m\\ 
\hspace{1cm} & Extinction in polar direction & E(B-V) & 1 - 4 (16 steps)  \\ \hline

\end{tabular}
\caption{A summary of the parameter choices made our \texttt{CIGALE} fitting. For the templates and parameters not shown in this table, we adopted default values from \texttt{CIGALE} }
\label{tab:table-one}
\end{table*}

Our analysis uses the \texttt{CIGALE} \citep{Burgarella_2005, boquien_cigale_2019} SED fitting code. \texttt{CIGALE} models the far-UV to radio spectrum of galaxies to estimate physical properties such as stellar mass, AGN luminosity, star formation rate, and dust attenuation. \texttt{CIGALE} composite models are built based on templates describing various aspects of the galactic emission, such as the star formation history (SFH), nebular emissions, dust emission and attenuation, and the AGN contribution. 

We fit the available photometry of the eight LRDs using SEDs built on a combination of two emission templates: stellar and AGN. We selected templates from the ones provided in \texttt{CIGALE} to vary only a small subset of parameters to focus our analysis on how the AGN fraction affects the SED fitting results. Table \ref{tab:table-one} lists the selected \texttt{CIGALE} templates and the parameters that were kept free, alongside the parameters where we adopted a different value from the default value. 
We note that we adopted the fiducial redshift values for each source, as provided by \cite{Perez_Gonzalez_2024}; hence, \texttt{CIGALE} does not fit redshifts.

For the star formation history, we use a standard delayed $\tau$ model with exponential burst: {\fontfamily{cmtt}\selectfont{sfhdelayed}}. We allow the $e$-folding time to vary from 0.1 to 5 Gyrs and the main stellar age from 0.1 Gyrs to the age of the Universe at the observed $z$. These values are typically used in the literature for similar analyses. For the simple stellar population, we use the {\fontfamily{cmtt}\selectfont{bc03}} module \citep{Bruzual_Charlot_2003}, with a \cite{Chabrier_2003} IMF. Although the module defaults to solar metallicity ($Z$ = 0.02), we find that changing this parameter had little effect on our modeling. To model for nebular emissions from the HII regions, we use the {\fontfamily{cmtt}\selectfont{nebular}} module from \cite{Inoue_2011}. We model the physics of the Lyman alpha emission using results from \citealt{Lin_2024}, which found a Lyman alpha escape fraction of 9\% from H$\alpha$ emitting galaxies observed by JWST between z = 4.9 - 6.3.

For dust attenuation, we employ the {\fontfamily{cmtt}\selectfont{dustatt\_modified\_starburst}} module based on the \citealt{Calzetti_2000} attenuation law, extended to lower wavelengths with \citealt{Leitherer_2002}. The Ly$\alpha$ result from \citealt{Lin_2024} employed a Small Magellanic Cloud dust law, yielding an E(B-V) color excess of 0.1, which we also adopt for our analysis. We employ the {\fontfamily{cmtt}\selectfont{dl2014}} module \cite{draine_andromedas_2014} for galactic dust emission. For spatially unresolved sources at high redshift, there is little information regarding the relative contributions of the torus and the interstellar medium to the total dust attenuation \citep{Hickox_review_2018}. For our analysis, we do not vary dust attenuation on a galactic scale; instead, we focus on the torus for simplicity.


For the AGN component, we employ the {\fontfamily{cmtt}\selectfont{skirtor2016}} module based on clumpy AGN models from \cite{Stalevski_12} and \cite{Stalevski_16}. {\fontfamily{cmtt}\selectfont{skirtor2016}} uses three parameters, $\tau$, $pl$, and $q$ to model the dust dynamics. The $\tau$ parameter sets the average edge on optical depth at 9.7 $\mu$m (possible values: 3,5,7,9,11), $pl$ is the power-law exponent that sets the radial gradient of the dust density (possible values: 0,0.5,1,1.5), and $q$ denotes the dust density gradient with polar angle (possible values: 0,0.5,1,1.5). Varying these parameters had little effect on the fit results but significantly increased the computation time for \texttt{CIGALE}. To reduce the compute time required, we fixed all three parameters. We adopt a modest opening angle of the torus at 40\degree. In {\fontfamily{cmtt}\selectfont{skirtor}}, the inclination angle determines the AGN type; therefore, we fix this parameter to either 30\degree for Type I templates or 70\degree for Type II templates. We do not vary the inclination for the two AGN types because the fit results are generally insensitive to different viewing angles.

The AGN fraction in {\fontfamily{cmtt}\selectfont{skirtor}} is defined as the ratio between the AGN luminosity and the total luminosity over a wavelength range defined by the user. We set the (rest-frame) wavelength range to 0.4 - 0.7 $\mu$m, to ensure that the AGN fraction is calculated over a range of wavelengths that is observed by JWST. We then fix the AGN fraction between 2\% and 99\% to determine the best set of parameters at a given AGN fraction. For the torus dust attenuation, we adopt a \citealt{Calzetti_2000} attenuation law, and we allow the torus reddening parameter to vary between 1 and 4, corresponding to an $A_v$ value of 3.63 and 14.52 respectively, as we expect this range of reddening values to better match the level of obscuration reported for JWST galaxies at $z \sim 6$. See Section 4 for a discussion on how fixing this parameter affects our SED fitting results.

\section{Results}
\label{sec:final}

In this Section, we first investigate how a range of AGN fraction values affects the goodness-of-fit descriptors. Then, we study how the inferred best-fit AGN fraction localizes the source in the \mmstar plane.

\subsection{AGN nature of LRD}

This analysis aims to constrain the nature of the LRDs while accounting for our lack of information on precisely how much the AGN contributes to the observed SED. To this end, we model our photometry using both Type I unobscured AGN and Type II obscured AGN templates. For all 19 galaxies in our analysis, there is a clear preference for the Type I AGN template, as it consistently returned better reduced $\chi^2$s than the obscured Type II template. The Type II templates typically yielded reduced $\chi^2$ values an order of magnitude larger than the Type I fits, often with such poor $\chi^2$ statistics that \texttt{CIGALE} could not return a best-fit model. Additionally, for galaxies where the AGN component is consistent with a type II template, AGN contribution to the rest optical wavelength emission is negligible, and so we do not explore these any further. Therefore, the remainder of this paper will focus on the results of the SED fitting using the Type I unobscured AGN template. 

In addition to this, we found the recovered \mmstar parameter space showed behavior that appears not to be physically motivated, such as little to no evolution in the estimated black hole mass as the AGN fraction increases, smaller AGN fraction corresponding to a larger black hole mass compared to higher AGN fractions in the same galaxy. We found the \texttt{CIGALE} code sensitive to the torus reddening parameter. Thus, our choice to allow the parameter to vary from 1 to 4 resulted in some extreme modeling configurations. However, we find that for all 19 galaxies, the best-fit AGN fraction occurred consistently at an $E(B-V)_{polar}$ value of 1. Therefore, we fix the torus reddening to a value of 1 and re-ran our analysis, the results of which will be discussed further below.

\subsection{Stellar-Dominated or AGN-Dominated?}

To constrain how the modeling of AGN contribution to the LRD SED affects the recovered stellar parameters, we perform a complete analysis of how a varying AGN fraction modifies the location of the galaxy in the \mmstar plane; we then use the reduced $\chi^{2}$ statistic provided by \texttt{CIGALE} to determine the best-fit parameters. 

The direct outputs from the SED fitting process outlined in Section \ref{sec:method} are the stellar mass of the host and the AGN bolometric luminosity. This latter parameter is then used to derive the mass of the central black hole by assuming that accretion occurs at the Eddington luminosity. 
Note that most of the lower-luminosity broad line AGN found at $z > 4$ by JWST thus far have a high mean Eddington ratio of close to unity (see the related discussion in, e.g., \citealt{Harikane_2023, Maiolino_2023_new, Pacucci_2024_evolution}). It is important to remark that lower values of the Eddington ratios would require a larger black hole mass, which would render our black holes even more overmassive. In principle, super-Eddington accretion rates may be possible in these high-$z$ environments given the wide availability of cold gas \citep{Power_2010, Pacucci_2020_separating}. However, for this analysis, we assume a maximum luminosity of approximately $L_{\rm Edd}$, as has been observed in other AGN populations across a range of redshifts (e.g., \citealt{Aird_2012, Fan_2022_review, Guangdong_2024}).

\begin{figure}
    \centering
    \includegraphics[width = \linewidth]{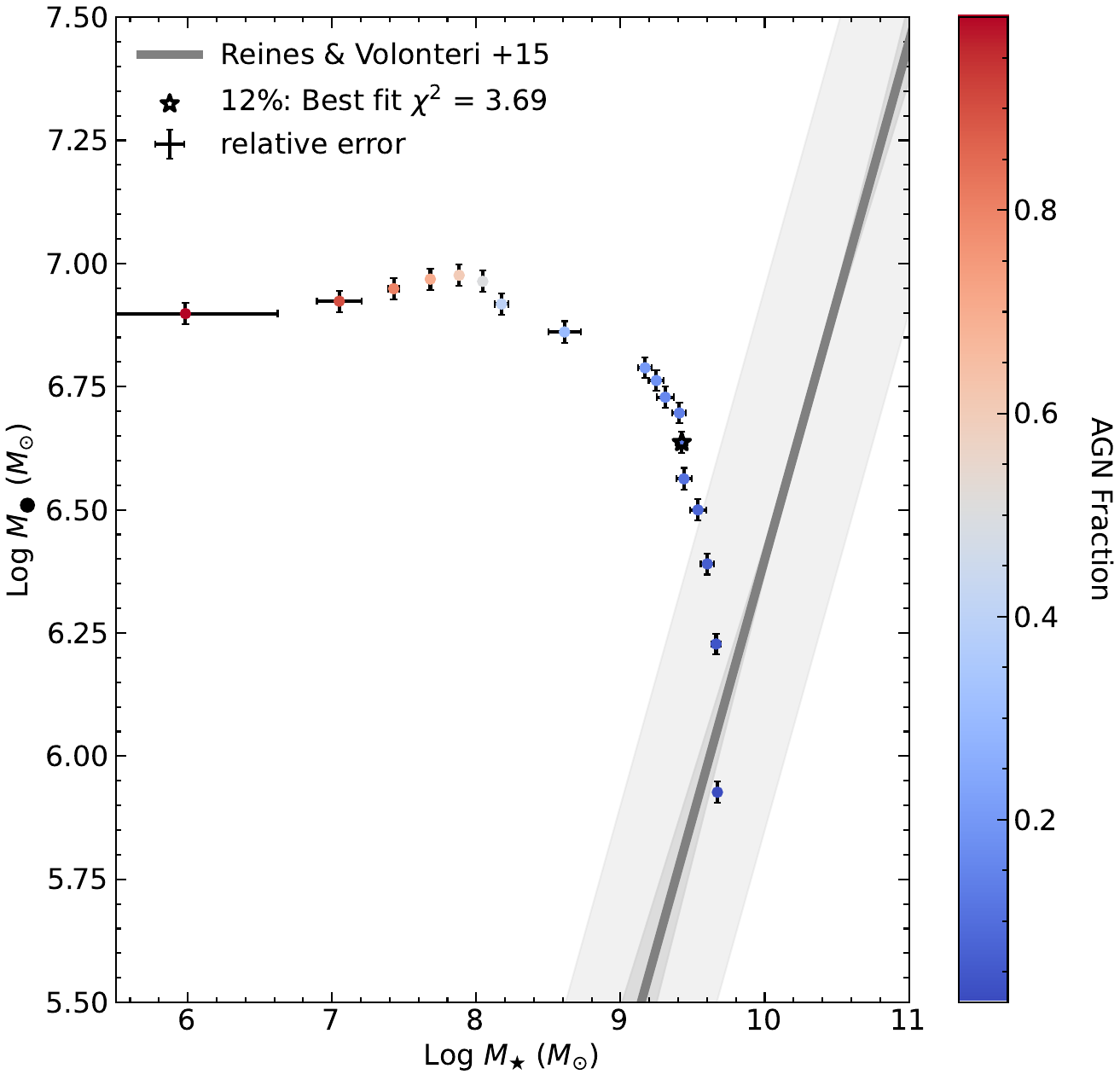}
    \caption{JADES-79803, study of how a varying AGN fraction (between 2\% and 99\%) changes the location of the source in the \mmstar plane. The grey line represents the \citealt{reines_relations_2015} local \mmstar relationship. We have also added the relative errors from \texttt{CIGALE}. Here, our best-fit value is 12\% AGN contribution (black star) with a reduced $\chi^2$ value of 3.69.}
    \label{fig:sample-size}
\end{figure}

This process provided estimates for the black hole and stellar masses for any given AGN fraction. An example of the \mmstar parameter space for one of our LRDs, JADES-79803, is shown in Figure \ref{fig:sample-size}; \textit{as the AGN fraction increases (i.e., higher AGN contribution to the overall SED), so does the inferred black hole mass, which then corresponds to a decrease in stellar mass}. This observed behavior matches our expectations based on how \texttt{CIGALE} uses composite models to construct SEDs.

Of the nineteen sources analyzed, we found one galaxy, JADES-187025, likely to be a star-forming galaxy with no AGN component. 11 out of the nineteen galaxies are best fit by an AGN fraction that is less than 40\% while the rest of the population had best-fit AGN fractions between 40\% and 70\%, suggesting that, in general, the SEDs of these systems are consistent with some, and in certain cases, strong AGN contribution. However, the precise value of the AGN fraction is not well constrained; further analysis, such as spectroscopic follow-up, would be required to confirm the contribution of the AGN to the total luminosity in these sources. Nevertheless, our results clearly indicate that AGN contribution is required to explain most LRDs in our sample, and that significant AGN contribution is required to explain the overmassive nature of the black holes hosted by LRDs.

\begin{figure*}

	\begin{minipage}[b]{0.6\columnwidth}
	\centering
	\subfloat[Best fit AGN fraction]{
		\includegraphics[clip, width=1\columnwidth]{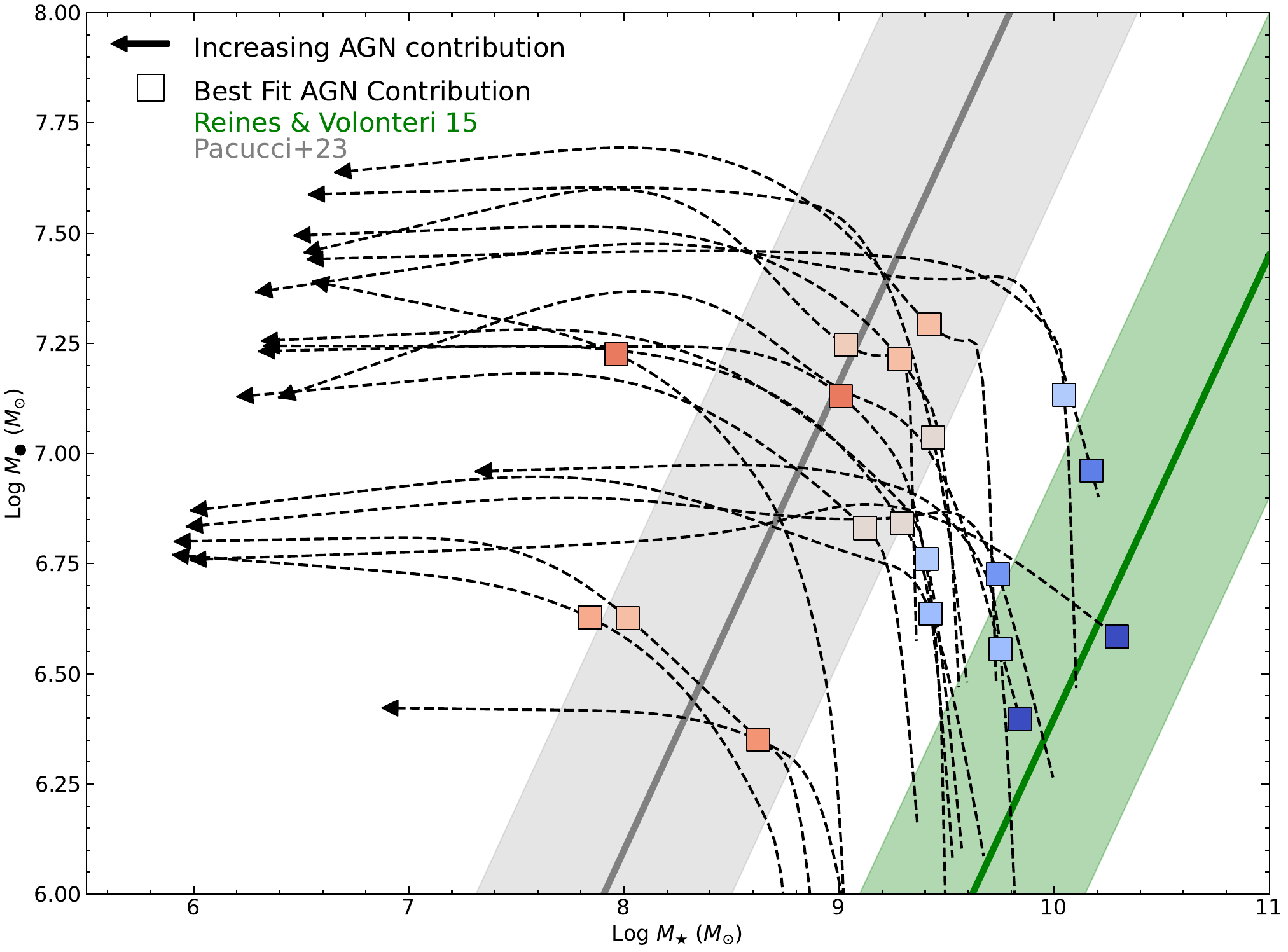}
		\label{best-fit}
	 } \hfill
	\subfloat[AGN fraction at the Balmer limit, H$_\infty$ ]{
		\includegraphics[clip, width=1\columnwidth]{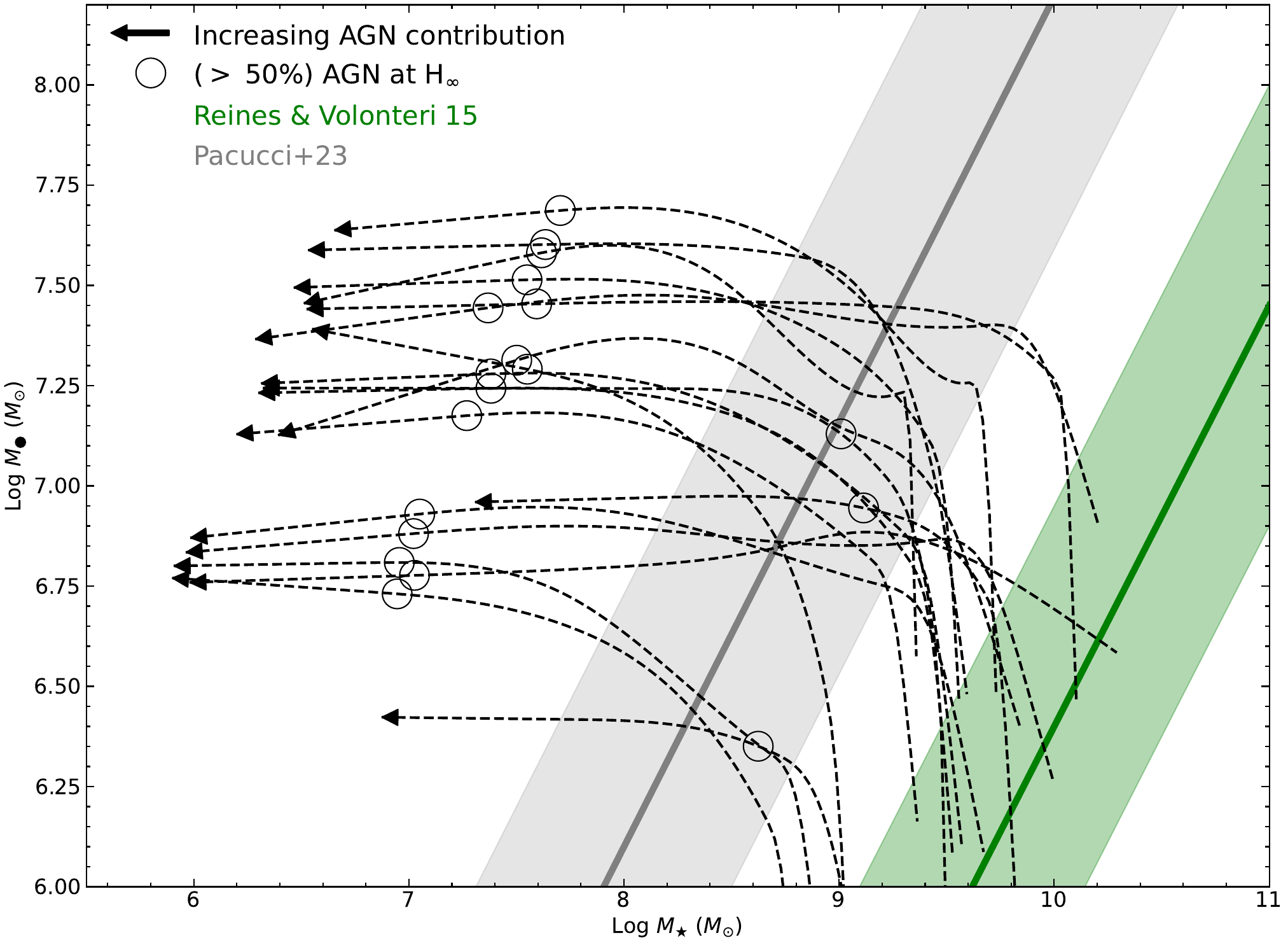}
		\label{Balmer-frac}
	}
    \end{minipage}
    \centering
    \subfloat[Galaxies with F277W - F444W $>$ 1.7]{
		\includegraphics[width=1.3\columnwidth]{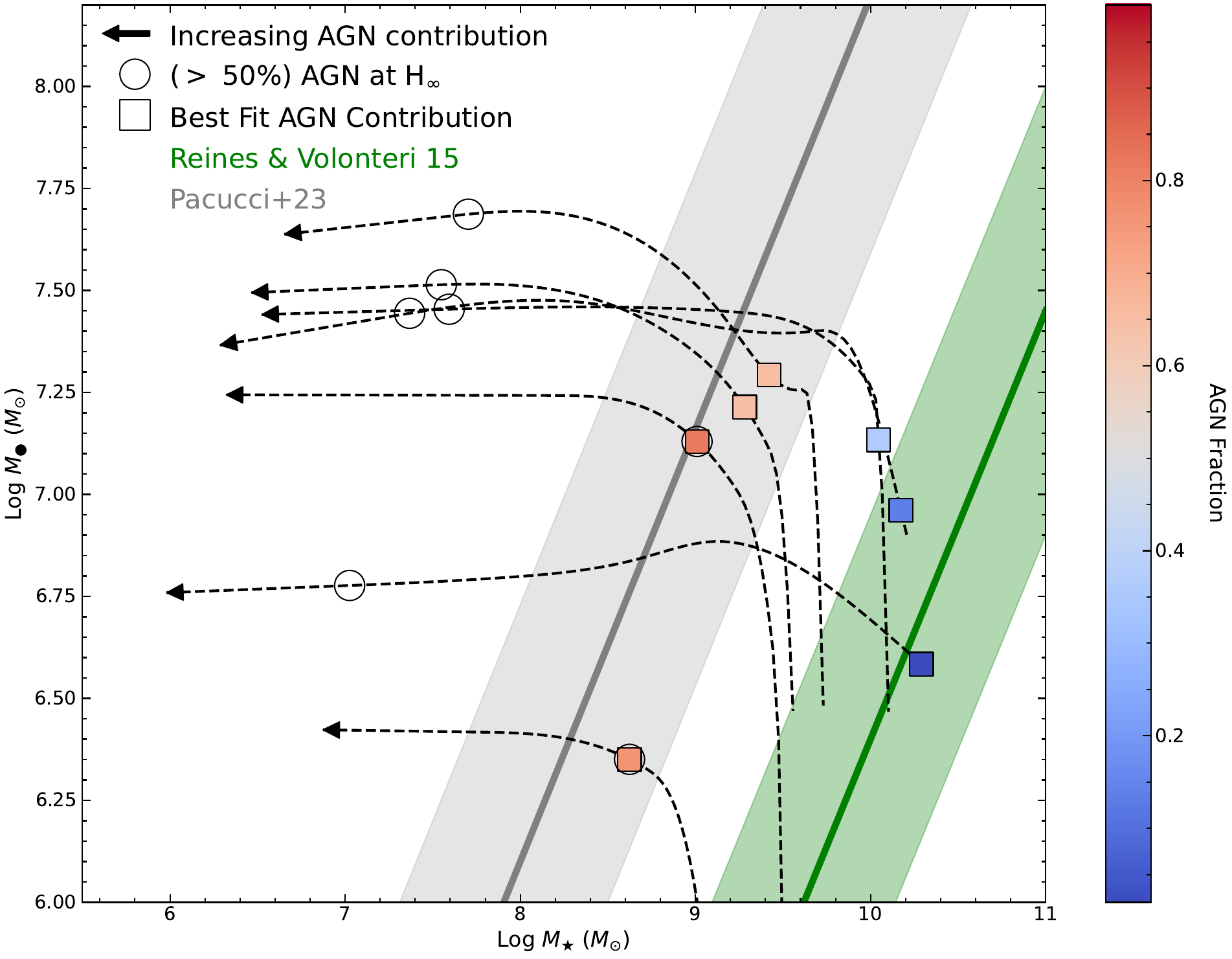}
		\label{reddest}
	}
    
\caption{The evolution of the \mmstar parameter space for the galaxies in our analysis. A.) The best-fit black hole and stellar masses of the 19 LRDs investigated are shown as square symbols for various fixed AGN fractions. The grey line represents the high-$z$ $M-M_\star$ scaling relation from \cite{pacucci_jwst_2023}, including its intrinsic scatter, while the green line represents the local scaling relation from \cite{reines_relations_2015}. The mean tracks illustrating how the black hole mass and stellar mass changes for each of the LRDs across varying AGN fractions are also displayed (for the full parameter space of black hole mass and stellar mass at each AGN fraction, see Figure \ref{fixed-red}). B.) Same as in (A) except that the circles represent where the AGN contribution at the blamer limit (H$_{\infty}$, 3645 \(\text{\AA}\)) exceeds 50\%. C.) The evolution of the reddest galaxies in our sample, defined as having F227W - F444W $>$ 1.7 \citep{greene_uncover_2023}. Circle symbols are as defined in (B) and the square symbols as defined in (A). \textit{Only two galaxies have best fit AGN fractions that coincide with the AGN contribution at the Balmer limit being above 50\%}}
\end{figure*}

\begin{figure}
    \centering
    \includegraphics[width=\linewidth]{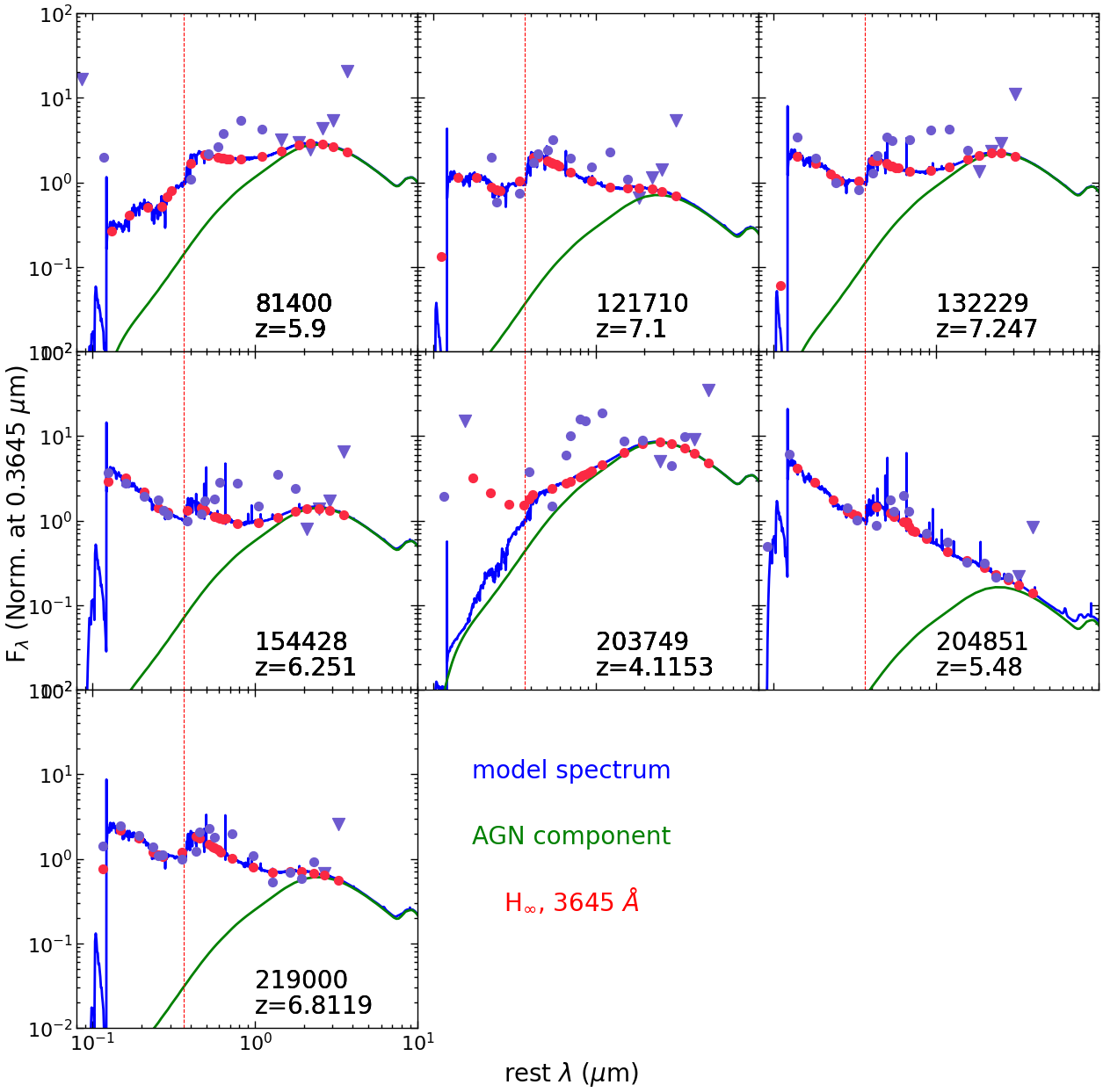}
    \caption{Rest wavelength F$_{\lambda}$ plot for the 7 objects with F277W - F444W $>$ 1.7. The blue component shows the \texttt{CIGALE} model continuum, while the green component shows only the AGN component. The red dashed line indicates the Balmer limit (H$_{\infty}$, $\lambda$ = 3645 $\AA$).}
    \label{fig:f_lambda}
\end{figure}

\begin{figure*}

    \centering
    \includegraphics[scale = 0.7 , angle=0]{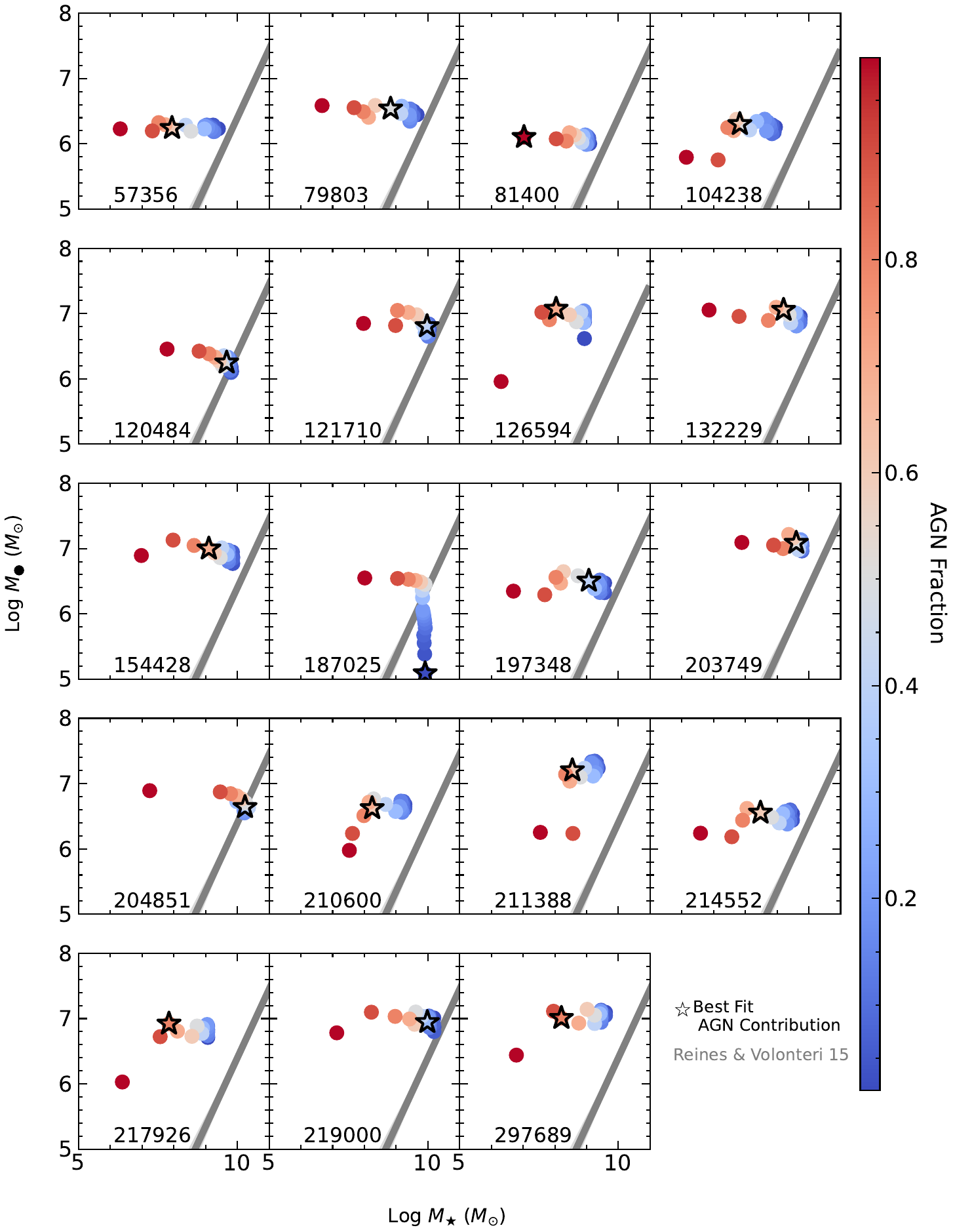}
    \caption{\mmstar parameter space for all 19 LRD, with sampled AGN reddening E(B-V)$_{polar}$ = 1 - 4.}
    \label{free-red}
\end{figure*}

\begin{figure*}

    \centering
    \includegraphics[scale = 0.7 , angle=0]{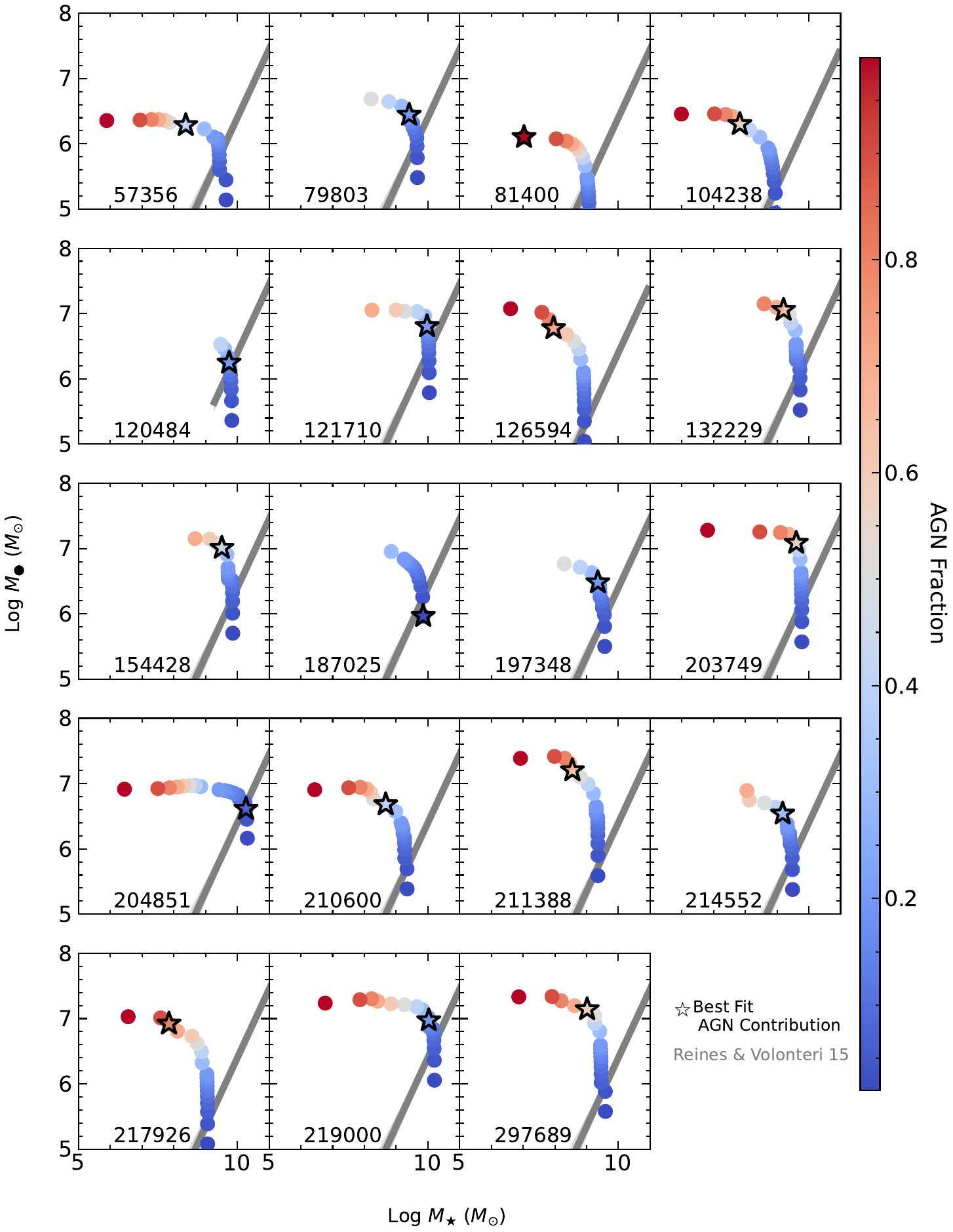}
    \caption{\mmstar parameter space for all 19 LRD, with fixed AGN reddening, E(B-V)$_{polar}$ = 1.}
    \label{fixed-red}
\end{figure*}

\subsection{Location in the \mmstar Plane}

The position of the 19 LRDs on the \mmstar plane for their best-fit models are shown by squares in Fig. \ref{best-fit}.
Generally, we find black hole masses ranging from $10^6$ to $10^{7.75}$ $M_{\odot}$.

We have also plotted the \cite{reines_relations_2015} $\Mblack-M_{\star}$ relation and the inferred high$z$ $\Mblack-M_{\star}$ relation from \cite{pacucci_jwst_2023}. 

We find that for galaxies where the best fit AGN contribution is below 20\%, the estimated black hole masses are in agreement with the local \citealt{reines_relations_2015} relation. For galaxies where the best fit AGN contribution is moderate (20\% $<$ frac$_{AGN}$ $<$ 40\%), the black hole masses are overmassive by $\gtrsim 1$ orders of magnitude compared to their local universe counterparts. For galaxies best fit by a significant AGN contribution (frac$_{AGN}$ $\geq$ 40\%), the estimated black hole masses are overmasive by $\gtrsim 2$ orders of magnitude (and up to $\sim 4$ in extreme cases). Hence, these specific LRDs with significant AGN contributions are best described by a scaling relation that favors overmassive black holes, such as the one described in \cite{pacucci_jwst_2023}.

Furthermore, we highlight that the presence of a black hole in these galaxies is supported by considerations of the limited amount of stellar mass that these hosts can contain, as they are very compact in size (see, e.g., \citealt{Pacucci_2024_evolution}). In fact, typical effective radii are in the order of $r_e \sim 150$ pc \citep{Baggen_2023}.
SED fitting executed with only stellar components (i.e., an AGN fraction of zero) would lead, in some cases, to average stellar densities that are larger, by factors $\sim 10-100$, than core densities in globular clusters (Pacucci et al., in prep.).

The same figure displays the mean track\footnote{We use the mean tracks here to illustrate the general behavior of the \mmstar plane, but it is not indicative of the true behavior. For the actual evolution of the \mmstar plane, we direct the reader to fig. \ref{fixed-red}} in the \mmstar plane that is due to a varying AGN fraction (in the range 2\% to 99\%) for each LRD. Interestingly, the tracks show that for an assumed modest AGN contribution ($\leq$ 20\%), the recovered AGN luminosity, and hence black hole masses are in agreement with the local \citealt{reines_relations_2015} relationship. However, assuming significant AGN contribution ($\gtrsim$ 40\%), SED fitting strongly suggests the presence of an overmassive black hole. Furthermore, for all 19 LRDs, the tracks evolve to the plot's left side, showing that as the AGN fraction increases, the objects become more overmassive on the \mmstar plane. We argue that this result shows that \textit{if the LRDs observed by JWST are AGN hosts, they are overmassive only if there is significant AGN contribution to the overall SED}.

Figure \ref{best-fit} shows two galaxies for which the best fit AGN fraction is negligible (frac$_{AGN}$ = 2\%), those two galaxies are Jades-187025 and Jades-204851.  
Jades-204851 as mentioned in \ref{sec:data}, is the only broad line detected galaxy in our sample, and so it is concerning that it is only best fit by an AGN contribution of 2\%. The surprisingly low best-fit AGN contribution was suggestive of the two galaxies being star-forming galaxies, instead of high redshift AGN hosts. For JADES-187025, this was indeed the case, as the SED fitting results for the stellar model return a better reduced $\chi^2$ compared to the composite (AGN + stellar) model, and so we conclude that it is a dusty star-forming galaxy. However, when we repeat the same analysis for JADES-204851, we find that the reduced $\chi^2$ for the stellar model is worse compared to the composite model. 

Interestingly, \citealt{Perez_Gonzalez_2024} also recovered a low AGN contribution of approximately 10\% when using a composite template of an unobscured QSO and a stellar template to fit the photometry for JADES-208451. Closer inspection of the \texttt{CIGALE} stellar model spectrum for JADES-204851 suggests that \texttt{CIGALE} could be underestimating the dust emission, which, if correctly modeled, could potentially lead to a much better SED fit for a stellar model. \texttt{CIGALE} modeling, much like the majority of the publicly available SED fitting codes, is based on an energy balance principle, where the absorbed light in the UV is remitted by the dust in the IR; this approach completely omits potential contributions from a hot dust component and is less reliable when fitting star-forming galaxies at high redshifts (z $>$ 4, \citealt{Haskell_2023}). And so it is plausible that JADES-204851 could be a high redshift star-forming galaxy, whose broad line emission is due to kinematics and not AGN activity \citep{Baggen_2024}.

\subsubsection{AGN contribution at the Balmer limit, H$_{\infty}$}

The spectral slope break at rest H$_{\infty}$ discussed in \citealt{Setton_2024} implies that an AGN model is inconsistent with explaining the spectroscopy of LRD. Here we assess the strength of the AGN contribution at the Balmer limit to determine the consistency of the composite model presented here under consideration of the spectral slope break. To constrain the AGN contribution at the Balmer limit, we define a new AGN fraction at rest H$_{\infty}$ (3645 \(\text{\AA}\)). We then define a threshold of 50\% to represent where the AGN contribution would impact the spectral slope at rest H$_{\infty}$. Figure \ref{Balmer-frac} shows the location along the \mmstar where the AGN contribution at the Balmer limit exceeds 50 percent. We find that for most of the galaxies analyzed here (16/19) an AGN contribution that exceed 50\% at the Balmer limit is consistent with a ($\gtrsim$ 90\%) mid-IR AGN fraction. Such a high mid-IR AGN fraction is well beyond the typical ranges we recover as best-fit AGN fractions, even in cases where the black hole masses are overmassive by up to a factor of $\gtrsim$ 4, see Figure \ref{best-fit}. \textit{Therefore, we conclude that for the population of LRD with spectral slope break at rest H$_{\infty}$, a significant mid-IR AGN ( $\leq$ 70\%) component that can produce an overmassive black hole result would preserve the spectral feature at rest H$_{\infty}$.}

Figure \ref{fig:f_lambda} shows the continuum from the \texttt{CIGALE} model for the seven galaxies with F277W - F444W $>$ 1.7. For three out of the seven galaxies, there is a prominent spectral break at the Balmer limit, similar to what was reported by \citealt{Setton_2024}. All three galaxies feature a best-fitting AGN fraction in the rest-frame mid-IR between 40 and 70 percent. This shows that a reddened AGN component can account for the extra red emission in the optical for LRD while maintaining the H$_{\infty}$ spectral break. We must emphasize, however, that this result does not rule out other possible sources for the optical red emission that is observed in LRD.

\subsubsection{Photometric selection of sources}

The LRD selection criteria adopted by \citealt{Perez_Gonzalez_2024} is largely consistent with other works (see \citealt{greene_uncover_2023, Kocevski_2023, barro_extremely_2023} ) in terms of color and compactness. The only discernible difference is the F277W - F444W magnitude criteria, where \citealt{Perez_Gonzalez_2024} uses a threshold of 1, other works used a threshold of 1.7. To test how this criterion affects our analysis, we restrict our sample only to include galaxies where the F277W - F444W is greater than 1.7. Adopting this color criterion reduced our galaxy catalog from nineteen to seven. Figure \ref{reddest} shows the \mmstar parameter space for the seven reddest galaxies in our catalog, while Table \ref{tab:z_reddest} reports their redshifts.

\begin{table}[htb]
\centering
\caption{Redshifts for sources with F277W - F444W $> 1.7$.}
\label{tab:z_reddest}
\begin{tabular}{cc}
\hline
\hline
\textbf{Source ID} & \textbf{Redshift} \\
\hline
81400   & 5.9 \\
121710  & 7.1 \\
132229  & 7.247 \\
154428  & 6.251 \\
203749  & 4.1153 \\
204851  & 5.48 \\
219000  & 6.8119 \\
\hline
\end{tabular}
\end{table}

Despite the smaller sample size, we find that our results still hold -- \textit{the mid-IR AGN contribution needed to produce overmassive black holes is still significant ($\gtrsim$ 40\%), while the threshold for the AGN emission to affect the spectral slope at rest H$_{\infty}$ is typically much higher than the best-fit mid-IR AGN fraction}.

\subsection{AGN reddening and the dependence of SED fitting on MIRI observations}
Our analysis identified two key factors that influence the behavior of the \mmstar parameter space: (1) the availability of MIRI photometry and (2) the modeling of polar reddening in the AGN torus.

\subsubsection{MIRI photometry}

One of the key results from the LRD analysis is that the characteristic ``v'' shape of the SEDs makes the SED fitting comparable for both reddened star-forming (SF) templates and composite SF+AGN templates \citep{barro_extremely_2023}. However, most LRDs analyzed by \citealt{barro_extremely_2023} had little to no MIRI photometry, with the deepest mid-IR data only reaching 0.77 $\mu$m (rest-frame). We believe this lack of mid-IR data plays a critical role in the ambiguity of LRD photometry. To better understand the impact of missing mid-IR photometry on the SED fitting of LRDs, we removed all MIRI data from our catalog and re-ran the analysis.

We found that the SED fitting results, as shown in Figure \ref{fig:sample-size}, are highly sensitive to mid-IR photometry. Without MIRI photometry, the fitting returned comparable $\chi^2$ statistics for both AGN and non-AGN templates. Furthermore, the absence of MIRI data led \texttt{CIGALE} to prefer high polar reddening, with the average AGN polar reddening being $\gtrsim$ 2 and typical values ranging between 4 and 20. This result underscores the importance of MIRI observations, as such extreme reddening would conflict with the significant portion of the LRD population observed to exhibit broad-line emission \citep{greene_uncover_2023}.

We also found that the depth of the included MIRI photometry affects the precision of \texttt{CIGALE} estimates of stellar masses. Specifically, we find that as the AGN contribution becomes maximal, the precision of the \texttt{CIGALE} stellar mass estimate worsens. However, we find that the precision becomes artificially enhanced as the included MIRI wavelengths become shorter. This result suggests that SED fitting analysis of LRD where the only considered AGN fraction is maximal could have some under-estimated errors on their stellar mass estimates, especially in cases where the MIRI photometry is not included.

The relevance of MIRI detections in our sample of LRDs warrants further discussion. Previous studies of LRDs (see, e.g., \citealt{Williams_2024, Akins_2024}) remarkably found very faint MIRI fluxes in the F770W and F1800W filters. Overall, this finding suggests a lack of hot dust emission. However, \cite{Li_2025} demonstrated that this faint emission in MIRI bands can be easily explained by an extended density distribution in the dust torus model. Other studies (see, e.g., \citealt{Wang_2024}) highlighted the presence of Balmer breaks as indicator of the prominence of stellar activity (however, see also \citealt{Inayoshi_Maiolino_2025}), and they also suggested that deeper and redder data are required to investigate the nature of the LRDs. The fact that this study focuses on LRDs with strong MIRI detections inherently limits our analysis to systems characterized by bright dust emission in the infrared.

\subsubsection{Polar reddening of AGN torus}
In addition to mid-IR photometry, we also found that the recovered stellar mass and AGN luminosity were highly sensitive to how we modeled the polar reddening, $E(B-V)_{polar}$. Here, we discuss how the modeling choices for the polar reddening, such as the attenuation law and the flexibility of the reddening parameter, E(B-V), affected our analysis results. 

Recent spectroscopic analysis of LRDs \citep{greene_uncover_2023, Maiolino_2023_new} have commonly modeled the AGN polar reddening using a Small Magellanic Cloud (SMC) dust law from \citealt{Pei_1993}. The attenuation law used in our analysis follows from the Balmer decrement analysis of LRDs from \citealt{Brooks_2024}, as mentioned in section \ref{sec:method}. To compare the effects of attenuation law on the SED fitting results, we re-ran our analysis using a SMC dust law. Using the SMC dust law resulted in a ($\sim$ 10\%) lower AGN luminosity compared to using the \citealt{Calzetti_2000} dust law. 

In addition to the choice of attenuation law, we also tested how the model choice for the reddening parameter would affect our analysis. Similar to our modeling with the \citealt{Calzetti_2000} dust law, we initially allowed \texttt{CIGALE} to sample $E(B-V)_{polar}$ values between 1 and 4. We find that even for the SMC dust law, the \mmstar parameter space performed poorly, and only when the torus reddening is fixed to a value of 1 do we then get a monotonically behaved \mmstar parameter space. This result suggests that the estimated stellar masses and black hole masses from photometric modeling of LRD are more sensitive to how the torus reddening is modeled and less the specific dust law used for modeling.

Figure \ref{free-red} shows the \mmstar space for all 19 galaxies modeled assuming a SMC dust law for the torus dust reddening, where the E(B-V)$_{polar}$ was sampled from 1 to 4. The behavior of the parameter space appears to be random, where, in some instances, the black hole mass at a lower AGN fraction is higher than at a higher AGN fraction. Additionally, at lower AGN fractions, we found the reddening to be systematically higher, ranging between 1.4 and 3, which would be inconsistent with any H$_\alpha$ emission being detected in these galaxies. Fixing the polar reddening value to 1 results in the \mmstar parameter space shown in Figure \ref{fixed-red}, where the relationship between the AGN fraction and the black hole mass is as expected.

\section{Discussion and Conclusions}
\label{sec:conclusions}

One of JWST's most surprising findings to date is the omnipresence of extremely red, compact objects in the high-$z$ Universe \citep{Kocevski_2024}. These faint, remarkably red sources have been selected in several surveys of different regions of the sky.

However, selecting these objects is challenging due to the degeneracy of the many processes that can cause a galaxy to look red. JWST photometry can look redder due to high equivalent width emission lines, which boosts the photometry in the JWST NIRCAM filters. Obscured AGNs could also contaminate the selection because the red power-law-like emission could result in red optical-to-infrared colors. Additionally, some LRDs have been re-classified as brown dwarfs \citep{nonino_23, Hainline_23, Kocevski_2024}. 

Hence, selecting high-$z$ galaxies based only on their red colors is associated with significant uncertainties. In addition, obtaining physical insights from fitting photometric SEDs can be challenging, as they can be modeled with components that are star-dominated or AGN-dominated while maintaining reasonable goodness-of-fit. 

However, despite these intrinsic limitations, our work has shown that rigorously probing the AGN vs stellar contribution to the SED can constrain our sample of LRDs as high-$z$ AGN hosts, as also suggested by previous studies \citep{greene_uncover_2023} of different samples. In our sample of 19 LRDs, we generally found that the best fit AGN contribution is between 40\% to 85\%. The inferred AGN bolometric luminosities show black holes that are significantly overmassive compared to the stellar mass of the host, based on expectations from the local \cite{reines_relations_2015} \mmstar. These sources better agree with the high-z \mmstar relation inferred by \cite{pacucci_jwst_2023} from JWST galaxies in the redshift range $4 < z < 7$.

We also found that JWST NIRCam photometry without MIRI coverage is not sensitive enough to properly constrain the AGN IR luminosity, resulting in excessively IR-luminous AGNs. Hence, based on our findings, we argue that deep MIRI coverage is essential to reveal the highly obscured population of the black hole population in the LRDs.

The fact that most of the black holes observed thus far by JWST at $z > 4$ are overmassive with respect to the host's stellar content is not necessarily surprising. Earlier works (e.g., \citealt{Volonteri_2023}) predicted that if JWST were to observe central black holes in high-$z$, smaller galaxies, they had to be overmassive. However, \cite{pacucci_jwst_2023} recently found that the entire dataset of galactic systems found thus far by JWST at $z > 4$ is so overmassive, and JWST's sensitivity in H$\alpha$ so deep, that their $M_\bullet/M_\star$ ratios cannot be explained by being extreme outliers from the local \mmstar relation. They seem to be, in fact, a different statistical population of objects, with black hole masses significantly overmassive. Intriguingly, these galactic systems seem to be located close to the local $M_{\bullet} - M_{dyn}$ and $M_{\bullet} - \sigma$ relations, suggesting that these two relationships are more fundamental than the \mmstar \citep{Maiolino_2023_new, Juodzbalis_2024}.
This population of sources is also X-ray weak, with several studies cross-matching samples of LRDs with X-ray catalogs and finding no or very few detections (see, e.g., \citealt{Ananna_2024, Yue_2024_X-ray, Maiolino_X-ray_2024}). One possibility to solve this tension would be to assume that the IR emission is overestimating the AGN luminosity, leading to lower black hole masses. Unusually large levels of obscuration \citep{Maiolino_X-ray_2024} or mildly super-Eddington accretion levels, leading to intrinsically X-ray weak SEDs \citep{Pacucci_Narayan_2024, Lambrides_2024}, could also explain the X-ray deficiency; the X-ray faintness of the LRDs would then require next-generation, high-resolution X-ray observatories, such as AXIS \citep{AXIS_2023}.

Why are these high-$z$ black holes so overmassive, and what would their subsequent cosmic evolution be? While these questions remain open and will require additional data and careful analysis, some partial answers may be already available. 

For example, some studies, based on cosmological simulations (e.g., \citealt{Scoggins_2023}) suggest that this may be the best evidence thus far available that these black holes formed at even higher redshifts ($z = 20-30)$ as heavy black hole seeds \citep{Agarwal_2013}. 
Larger samples of red and compact objects at high redshift, as well as a better understanding of the uncertainties associated with their black hole and stellar mass measurements, will be vital to confirm the very existence of this remarkable population of overmassive black holes and their broader cosmological implications.

\vspace{10pt}
\noindent \textit{Acknowledgments:} 
We thank Pablo Perez-Gonzalez for providing the photometry of the 19 sources investigated here. E.D. acknowledges support from a Dartmouth Fellowship.
F.P. acknowledges support from a Clay Fellowship administered by the Smithsonian Astrophysical Observatory. R.C.H. acknowledges support from NASA through Astrophysics Data Analysis Program grant 80NSSC23K0485. This work was also supported by the Black Hole Initiative at Harvard University, which is funded by grants from the John Templeton Foundation and the Gordon and Betty Moore Foundation. 


\bibliography{ms}
\bibliographystyle{aasjournal}



\end{document}